\documentclass[sigplan,screen]{acmart}

\AtBeginDocument{%
  }

\setcopyright{acmcopyright}
\copyrightyear{2024}
\acmYear{2024}
\acmDOI{XXXXXXX.XXXXXXX}

\acmConference[Conference acronym 'XX]{Make sure to enter the correct
  conference title from your rights confirmation emai}{October 03--05,
  2024}{Woodstock, NY}
\acmPrice{15.00}
\acmISBN{978-1-4503-XXXX-X/18/06}

\usepackage{caption}
\usepackage{graphicx}
\usepackage{float} 
\usepackage{subcaption}
\usepackage{threeparttable}
\usepackage{mdframed}
\usepackage{multirow}
\usepackage{makecell}
\usepackage{threeparttable}
\usepackage{tabularx}
\usepackage{hyperref}

\begin{document}

\title{LLM-SmartAudit: Advanced Smart Contract Vulnerability Detection}


\author{Zhiyuan Wei}
\affiliation{%
  \institution{Beijing Institute of Technology}
  \city{Beijing}
  \country{China}}
\email{weizhiyuan@bit.edu.cn}

\author{Jing Sun}
\affiliation{%
  \institution{University of Auckland}
  \city{Auckland}
  \country{New Zealand}
}
\email{jing.sun@auckland.ac.nz}

\author{Zijian Zhang}
\affiliation{%
 \institution{Beijing Institute of Technology}
 \city{Beijing}
 \country{China}}
 \email{zhangzijian@bit.edu.cn}

\author{Xianhao Zhang}
\affiliation{%
  \institution{Beijing Institute of Technology}
  \city{Beijing}
  \country{China}}
  \email{zhangxh@bit.edu.cn}

\author{Meng Li}
\affiliation{%
  \institution{Hefei University Of Technology}
  \city{Hefei}
  \state{Anhui}
  \country{China}}
  \email{mengli@hfut.edu.cn}

\author{Zhe Hou}
\affiliation{%
  \institution{Griffith University}
  \state{Queensland}
  \country{Australia}}




\renewcommand{\shortauthors}{Wei et al.}

\begin{abstract}
  The immutable nature of blockchain technology, while revolutionary, introduces significant security challenges, particularly in smart contracts. These security issues can lead to substantial financial losses. Current tools and approaches often focus on specific types of vulnerabilities. However, a comprehensive tool capable of detecting a wide range of vulnerabilities with high accuracy is lacking. 
  This paper introduces LLM-SmartAudit, a novel framework leveraging the advanced capabilities of Large Language Models (LLMs) to detect and analyze vulnerabilities in smart contracts. Using a multi-agent conversational approach, LLM-SmartAudit employs a collaborative system with specialized agents to enhance the audit process. 
  To evaluate the effectiveness of LLM-SmartAudit, we compiled two distinct datasets: a labeled dataset for benchmarking against traditional tools and a real-world dataset for assessing practical applications. Experimental results indicate that our solution outperforms all traditional smart contract auditing tools, offering higher accuracy and greater efficiency. Furthermore, our framework can detect complex logic vulnerabilities that traditional tools have previously overlooked. Our findings demonstrate that leveraging LLM agents provides a highly effective method for automated smart contract auditing.
\end{abstract}



\received{20 February 2007}
\received[revised]{12 March 2009}
\received[accepted]{5 June 2009}

\maketitle

\section{Introduction}
Recent advances in smart contracts have marked significant progress in areas such as security, finance, and governance. These are self-executing contracts with terms mutually agreed upon by participants, enacted through predefined actions. However, the immutable nature of blockchain technology inherently makes smart contracts more susceptible to software attacks. Over the past years, there have been numerous high-profile vulnerabilities and exploits in smart contracts, such as the DAO attack \cite{atzei2017survey}. Consequently, the development of secure smart contracts continues to pose significant challenges.

Research in Large Language Models (LLMs) has made significant advancements in fields such as Natural Language Processing \cite{danilevsky2020survey}, Computer Vision \cite{ramesh2022hierarchical}, Code Generation \cite{dong2023self}, and various AI tasks \cite{shen2024hugginggpt, qian2023communicative}. A survey \cite{zhang2023survey} indicates a significant increase in LLMs adoption over the past five years, accompanied by a rapid rise in software engineering. LLM tools are primarily categorized into commercial products and open-source initiatives. State-of-the-art (SOTA) commercial LLMs include GPT \cite{OpenAI}, Claude \cite{claude2024}, and Gemini \cite{team2024gemma} models, while prominent open-source projects feature Llama \cite{touvron2023llama}, Mixtral \cite{jiang2024mixtral}, and GPT-NeoX \cite{black2022gpt}. Both commercial and open-source LLMs demonstrate significant potential in handling complex tasks. GPT, one of the pioneering commercial LLMs, has exhibited remarkable proficiency in natural language understanding, context processing, and code comprehension, outperforming numerous traditional methods \cite{xu2022systematic}. 

LLMs, trained on vast text data, excel at predicting likely token sequences based on input. Unlike traditional systems with fixed rules or vulnerability databases, LLMs generate probabilistic outputs. While powerful, this approach can sometimes lead to inaccuracies in vulnerability identification \cite{azamfirei2023large, kang2023large, chen2023diversevul}. This paper investigate the question: `\textbf{Can multi-agent conversations enhance LLMs' capabilities in detecting smart contract vulnerabilities?}'.

We propose a virtual chat-powered smart contract audit framework utilizing a multi-agent conversation system. This approach is based on three key rationales. 
%
Firstly, LLMs' ability to incorporate feedback and engage in cooperative conversations enables a dynamic, interactive, and iterative approach to identifying and addressing smart contract vulnerabilities. Secondly, the collaboration of agents with specialized knowledge and analytical skills enhances the factual accuracy and reasoning precision of the audit process \cite{salewski2023context}. Lastly, multi-agent interaction mitigates the `degeneration-of-thought' issue common in single-agent self-reflection processes \cite{liang2023encouraging}. This approach allows agents to challenge and complement each other's viewpoints, resulting in a more balanced and comprehensive analysis. 
Supporting this, Wu et al. \cite{wu2023autogen} demonstrated that multi-agent systems like AutoGen and CAMEL \cite{li2023camel} outperform single-agent systems such as AutoGPT \cite{yang2023auto} and LangChain Agents \cite{LangChain2023} across various applications. These findings underscore the potential effectiveness of multi-agent systems in conducting smart contract audits.

We developed LLM-SmartAudit, an innovative system designed to automate smart contract security analysis. LLM-SmartAudit employs two strategies to enhance LLMs' vulnerability detection capabilities: Broad Analysis (BA) and Targeted Analysis (TA). 
Experimental results demonstrate that LLM-SmartAudit significantly outperforms leading traditional tools in detecting vulnerabilities. This research provides several notable advancements in smart contract security:
\begin{itemize}
\item \textbf{The Power of Multiple LLM Agents}: Our research reveals that each LLM Agent, each assembled with the specific capabilities and roles, specialize in distinct areas of the security auditing, including contract code analysis, vulnerability identification, and Comprehresive Report. These specialized agents, guided by step-by-step instructions, perform in-depth analysis within their respective domains, yielding more accurate and comprehensive results. Furthermore, LLM Agents collaborate seamlessly, exchanging data and insights to provide a holistic view of the security landscape.

\item \textbf{Innovative Operational Strategies and System Enhancements}: Our research demonstrates that the operational strategies in the LLM-SmartAudit system have enhanced the ability of vulnerability predictions. Specifically, the TA mode is highly effective at detecting known vulnerability types, while the BA mode excels at identifying a broader spectrum of potential vulnerabilities. Together, these modes significantly reduce the variability and unpredictability often associated with LLM outputs, thereby enhancing the reliability of the analyses.

\item  \textbf{Benchmarking and Evaluation}: To evaluate our system, we leveraged two datasets: Labeled dataset and Real-world dataset. The Labeled dataset serves as a benchmark, enabling direct comparison between LLM-SmartAudit and conventional tools employing various analysis techniques. The Real-world dataset, derived from the reputable Code4rena contest, comprises 6,454 contracts across 102 smart contract projects. This dataset serves as a practical testbed to demonstrate our system's effectiveness in real-life conditions. Evaluations on both datasets demonstrate that LLM-SmartAudit not only excels in detecting common vulnerabilities but also outperforms existing solutions in identifying complex logic vulnerabilities.
\end{itemize}

The rest of the paper is organized as follows. Section \ref{sec:backgrounds} presents the current challenges in smart contract analysis and surveys the landscape of existing tools.  Section \ref{sec:gpt-scas} details the architecture and operational mechanism of LLM-SmartAudit. Section \ref{sec:experiments} presents a comprehensive evaluation of LLM-SmartAudit, detailing the standardized datasets, evaluation criteria, experimental results, and related works.
Section \ref{sec:discussion} discusses the findings and addresses potential threats to validity. Finally, Section \ref{sec:conclusion} concludes the paper, summarizing the contributions and offering insights into the future of smart contract analysis tools.

\section{Backgrounds}
\label{sec:backgrounds}

\subsection{Smart Contract Security}
Smart contracts, self-executing agreements encoded in software, facilitate the management and execution of digital assets within blockchain networks \cite{luu2016making}. These contracts not only define rules and penalties in a similar way to traditional contracts but also automatically enforce these obligations. Although Nick Szabo first proposed the concept in 1994 \cite{szabo1996smart}, smart contracts become practically implementable with Ethereum's launch in 2015. A recent survey \cite{zhou2023sok} reveals a rapid increase in the number of Solidity contracts over the past five years. This growth reflects the expanding range of smart contract applications across sectors such as decentralized finance (DeFi) \cite{berg2022empirical}, insurance, and lending platforms. Notably, the DeFi sector has experienced significant growth, with its peak total value locked (TVL) reaching 179 billion USD on November 9, 2021. Of this, Ethereum accounts for 108.9 billion USD, representing 60\% of the total DeFi TVL \cite{DeFillama2023}.

The substantial asset value managed by smart contracts underscores their critical security importance. However, a defining characteristic of Solidity smart contracts is their post-deployment immutability on the Ethereum network, presenting significant security management challenges. Unlike traditional software, where patches or updates can rectify bugs or flaws, smart contracts lack this flexibility. Consequently, vulnerabilities discovered post-deployment remain unfixable in the existing contract, potentially leading to substantial financial losses if exploited by malicious actors. According to Zhou et al. \cite{zhou2023sok}, smart contracts have been the target of numerous high-profile attacks, resulting in losses exceeding 3.24 billion USD from April 2018 to April 2022.

\subsection{Automated Security Analysis}
With the rise in security incidents and high-profile attacks, diverse smart contract analysis tools have been developed. These tools are designed to systematically detect vulnerabilities, enforce best practices, and identify potential security risks inherent in smart contracts. By leveraging these tools, developers can proactively mitigate issues before malicious exploitation, significantly reducing security breach risks and ensuring contract execution integrity. These tools employ advanced techniques such as formal verification, symbolic execution, intermediate representation (IR), and machine learning to enhance their effectiveness \cite{tolmach2021survey,chen2020survey}.

Despite these advancements, substantial challenges persist in smart contract security analysis. A primary concern is the complexity and diversity of vulnerabilities, making it difficult for any single tool to be universally effective. Each tool has its own strengths and limitations. For instance, tools relying on formal verification excel at ensuring contracts adhere to specified requirements but may fall short in detecting security flaws like reentrancy or gas limit issues. Complex logic vulnerabilities still necessitate human auditors \cite{code4rena2024}, introducing additional challenges. However, the fees charged by traditional smart contract auditing firms are prohibitively high. Basic audits from firms like CertiK start at 500 USD, while more reputable companies such as Trail of Bits charge between 5,000 USD to 15,000 USD as a starting point \cite{Mateusz2024}. 

\subsection{LLMs for Vulnerability Prediction}
LLMs, such as GPT and Claude, are a specific type of generative AI focused on text generation \cite{epstein2023art}. These models are termed "large" due to their vast number of parameters, enabling them to comprehend and produce human language with remarkable coherence and contextual appropriateness. Pre-trained on diverse internet-based text sources, they can produce text that often mirrors the quality and style of human writing. LLMs have demonstrated the ability to grasp grammatical structures, word meanings, and basic logical reasoning in human languages.

LLMs have demonstrated excellence in specific downstream tasks, including code completion, code analysis, and vulnerability detection. By leveraging their code comprehension and generation capabilities, these models can identify vulnerabilities, verify compliance, and assess logical correctness. Their effectiveness is further enhanced through advanced prompting techniques like chain-of-thought (CoT) or few-shot. Chen et al. \cite{chen2023diversevul} have proven that LLMs (GPT-2, T5), trained with a high-quality dataset consisting of 18,945 vulnerable C/C++ functions, outperform other machine learning methods, including Graph Neural Networks, in vulnerability prediction.  Khare et al. \cite{khare2023understanding} found that proper prompting strategies that involve step-by-step analysis significantly improve the performance of LLMs (GPT, CodeLlama) in detecting security vulnerabilities in programming languages such as Java and C/C++.

\begin{figure*}[ht]
  \centering
  \includegraphics[width=1.0\linewidth]{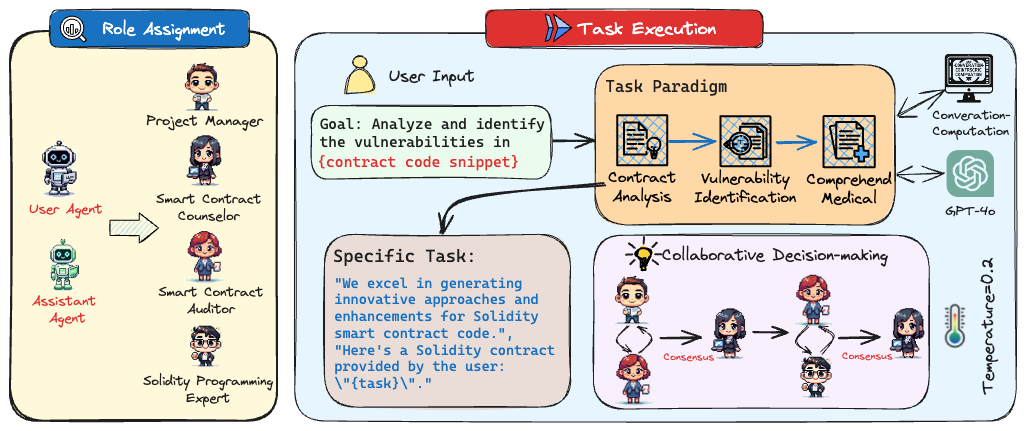}
  \caption{Muliti-agent Conversation Framework}
  \label{framework}
\end{figure*}

\section{LLM-SmartAudit System}
\label{sec:gpt-scas}

This section introduces LLM-SmartAudit\footnote{\url{https://github.com/LLMAudit/LLMSmartAuditTool}}, an innovative system designed to identify potential smart contract vulnerabilities. LLM-SmartAudit employs a multi-agent conversation approach, facilitating an interactive audit process. The system conceptualizes the analysis of smart contract codes as a specific task, autonomously executed through conversations among specialized agents. These conversations are structured as assistant-user cooperative scenarios, fostering a collaborative approach to achieve accurate and comprehensive smart contract audits.

\subsection{Framework}
The LLM-SmartAudit framework is found on two core principles: role specialization and action execution, as illustrated in Figure \ref{framework}. Role specialization ensures that each agent focuses on specific tasks, maintaining an efficient conversation flow. Action execution streamlines the agents' collaborative efforts, enhancing the overall efficiency and coherence of the audit process.

\subsubsection{Role Specialization}
The framework adopts a role-playing methodology to define and specialize the function of each agent within the audit process. By utilizing the inception prompting technique \cite{li2023camel}, the system enables agents to effectively assume and fulfill their designated roles. Agents are assigned specific capabilities and roles, either through repurposing existing agents or extending their functionalities. These roles encompass sending messages and receiving information from other agents, facilitating the initiation and continuation of inter-agent conversations.

The framework deploys a combination of an LLM-powered assistant agent and a user agent in an assistant-user cooperative scenario. The assistant agent, powered by LLMs \cite{OpenAI}, generates solutions and communicates these to the user agent. The user agent then executes the assistant’s recommendations, providing feedback to the assistant agent.

The system assigns specialized roles to agents, including Project Manager, Smart Contract Counselor, Auditor, and Solidity Programming Expert. These agents dynamically alternate between assistant and user roles in various audit scenarios, providing flexibility and depth to the audit process. The user agent primarily functions as a task planner, strategizing the audit approach and defining objectives. Conversely, the assistant agent serves as a task executor, performing detailed analyses and generating insights based on its specialized role.

\subsubsection{Action Execution}
After role specialization, assistant and user agents collaborate in an instruction-following manner to complete the assigned tasks.  Human users initiate the action execution by inputting contract codes into the system. The system segments this code into a task queue, which is systematically processed through three distinct phases: Contract Analysis, Vulnerability Identification, and Comprehresive Report. The Contract Analysis phase involves preliminary assessments of the contract's purpose and structure. During Vulnerability Identification phase,  agents collaboratively identify and describe potential security weaknesses. Finally, the Comprehresive Report phase generates a comprehensive audit report.

At each decision point within these subtasks, assistant and user agents engage in structured conversation, combining their respective insights to make informed decisions. The system employs a collaborative decision-making strategy, ensuring that the unique capabilities of each agent are leveraged to their fullest potential. This strategy ranscends mere task sharing, fostering a collaborative environment where LLM capabilities and human expertise are integrated for optimal outcomes.

\begin{figure*}[t]
    \centering
    \includegraphics[width=0.8\linewidth]{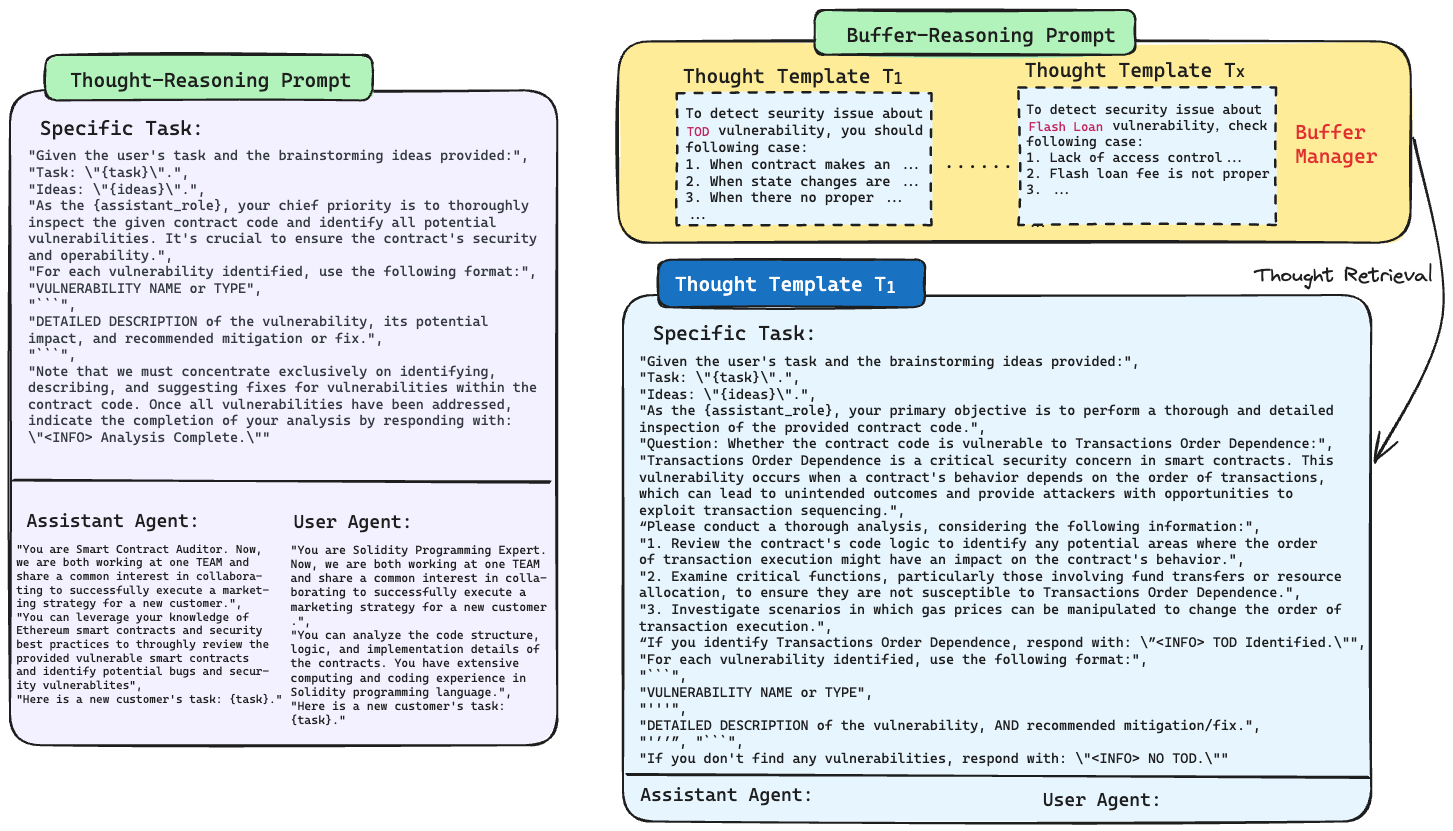}
    \caption{Thought-Reasoning and Buffer-Reasoning Prompting Strategies}
    \label{fig:task_queue}
\end{figure*}

\subsection{Task Queue}
\label{taskqueues}
In the LLM-SmartAudit framework, the task queue plays a vital role, guiding the system through a series of structured subtasks. This mechanism is essential for maintaining an efficient and coherent audit process.

\subsubsection{Structured Prompting for Task Specialization}  
Prompt engineering is a crucial element in achieving role specialization and efficient task execution within our framework. To enable this, we employ inception prompting \cite{li2023camel}, a technique that defines the roles, tasks, and responsibilities of both assistant and user agents. Each inception prompt includes three essential components: the specified task prompt, assistant agent prompt, and user agent prompt. This structured approach transforms initial concepts into specific, actionable tasks that harness the unique capabilities of each agent. Building on this foundation, our framework incorporates two strategies: Thought-Reasoning and Buffer-Reasoning.

\textbf{Thought-Reasoning prompt} originates from the ReAct framework \cite{yao2022react}, which effectively combines reasoning and acting in language models to handle complex tasks with adaptive thought and action. Our approach builds on this foundation, extending ReAct’s synergy between reasoning and action to ensure that the model not only processes information sequentially but also actively interacts with external data sources as needed. Unlike single-query (Few-shot or CoT) to model, which may lead to surface-level analyses, this method continuously verifies and refines reasoning through targeted interactions, enhancing depth and accuracy in complex task-solving.

Figure \ref{fig:task_queue} illustrates the Thought-Reasoning prompt template, adapted specifically for broad-spectrum vulnerability analysis. The \textit{specified task} sets out the main goal—identifying various vulnerabilities within GPT's knowledge base—by clearly defining expected outputs, including vulnerability types and detailed descriptions, alongside constraints to mitigate potential undesirable behaviors. The \textit{assistant agent} and \textit{user agent} provide structured guidance on roles and tasks, fostering collaboration between specialized agents to systematically uncover vulnerabilities in smart contracts.

\textbf{Buffer-Reasoning prompt}, derived from the foundational Buffer of Thoughts (BoT) \cite{yang2024buffer} concept, enhances the model's task comprehension by drawing upon high-level, context-specific thought templates, which is particularly crucial for complex domains. While ReAct is effective for many tasks, it often falls short in highly specialized areas, as LLMs generally lack domain-specific knowledge stores \cite{touvron2023llama}. LLMs typically lack domain-specific knowledge vault, limiting their performance in specialized fields. Buffer-Reasoning prompt addresses this limitation by retrieving relevant thought templates from prior problem-solving processes, supporting in-context learning and bolstering the model’s understanding. Unlike ReAct, Buffer-Reasoning prompt inherently guides LLMs to engage in deep, step-by-step reasoning required for tackling complex tasks. Building on this foundational concept, LLM-SmartAudit utilizes an adapted Buffer-Reasoning prompt approach. This method combines thought-augmented reasoning with adaptive instantiation, prompting LLMs to integrate contextual information and systematically analyze the problem.

Figure \ref{fig:task_queue} presents an example of the \textit{specified task prompt} for Buffer-Reasoning prompt, tailored for focused and detailed analysis of smart contracts. This prompt establishes a clear objective for the agents, concentrating on identifying Transactions Order Dependence (TOD) vulnerabilities. It directs agents to examine the contract's logic and critical functions, particularly those involving fund transfers, resource allocation, or gas price manipulation, for TOD vulnerabilities. If a TOD vulnerability is identified, the assistant agent must provide a detailed description of the vulnerability and its potential impact. Conversely, if no TOD vulnerability is found, the agent outputs ``NO TOD".

\subsubsection{Execution Mode}
The LLM-SmartAudit task queue comprises three primary subtasks: Contract Analysis, Vulnerabilities Identification, and Comprehresive Report. Each subtask employs task-oriented role-playing, involving two distinct roles collaborating to achieve specific objectives. Initially, the Project Manager establishes the primary goal for the team and collaborate with the Smart Contract Auditor to assess the contract's purpose and structure. The Smart Contract Counselor then reviews and summarizes these initial findings. This preliminary analysis, along with the smart contract codes, is then forwarded to the next phase. 

\begin{figure*}[t]
    \centering
    \includegraphics[width=0.8\linewidth]{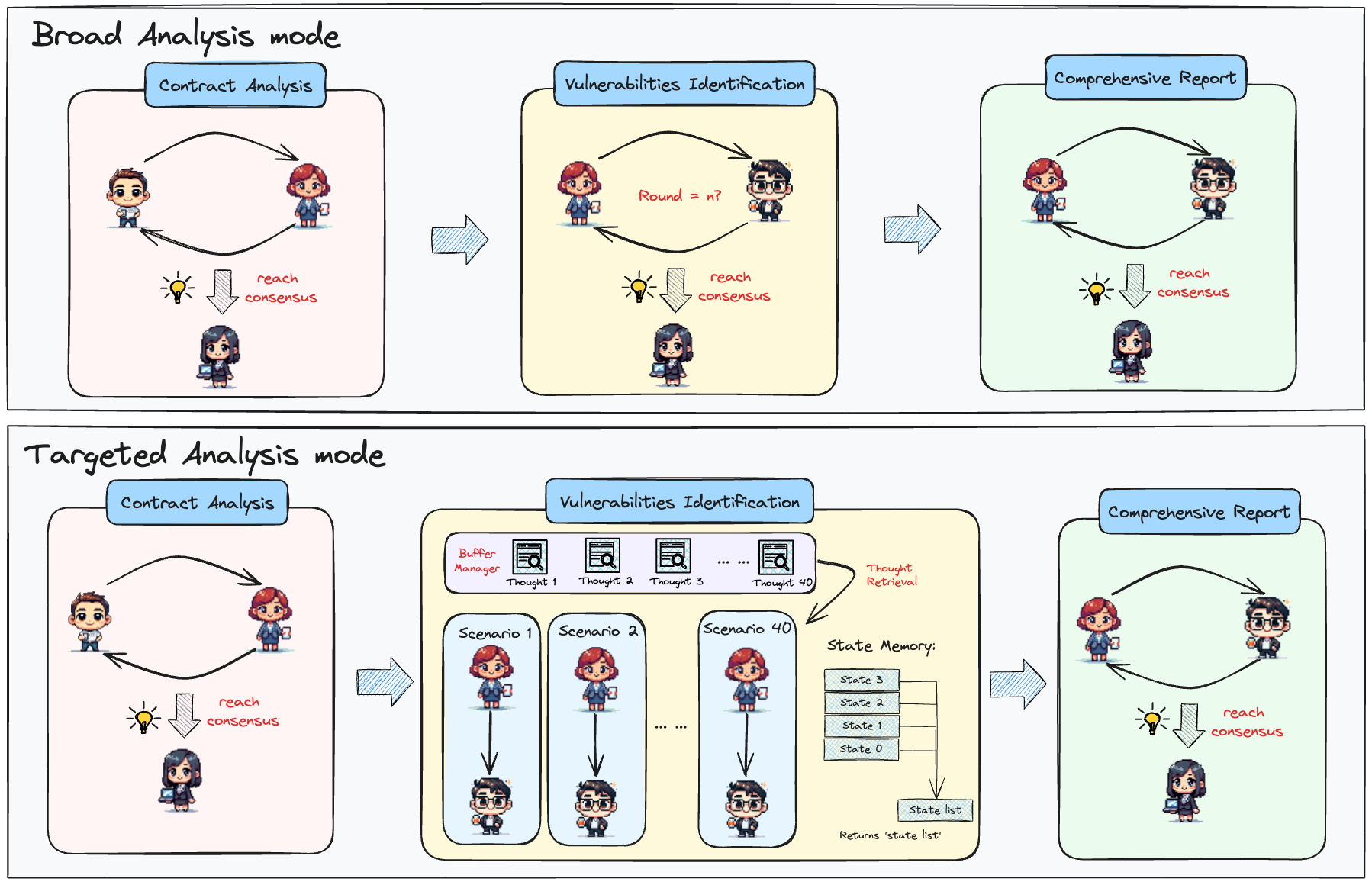}
    \caption{Task Queue in BA mode and TA mode}
    \label{task_queue}
\end{figure*}

In the next subtask, the Smart Contract Auditor and the Solidity Programming Expert work together to identify security weaknesses within the contract. In the final subtask, the Smart Contract Auditor and the Solidity Programming Expert jointly formulate a comprehensive analysis report, detailing all identified vulnerabilities and their potential impacts.

To assess the effectiveness of this workflow, LLM-SmartAudit operates in two distinct modes, illustrated in Figure \ref{task_queue}: Broad Analysis (BA) and Targeted Analysis (TA).

\textbf{BA mode}, based on the Thought-Reasoning approach, focuses on a thorough, comprehensive examination of smart contracts. In this mode, the expertise of the Smart Contract Auditor and Solidity Programming Expert is augmented with strategic guidance from the Counselor, leveraging the model’s capacity to identify a wide range of potential vulnerabilities. BA mode is particularly effective for its adaptability in handling various types of vulnerability detection tasks. To prevent potential impasses, where agents might struggle to reach a consensus on certain vulnerability assessments, LLM-SmartAudit restricts the maximum number of interaction rounds to a predefined number, $n$.

\textbf{TA mode}, based on the Buffer-Reasoning prompt, employs a scenario-specific approach for vulnerability detection. It divides the Vulnerability Identification phase into 40 targeted scenarios, each focused on a specific vulnerability type (as listed in our repository). This mode harnesses the collaborative efforts of the Smart Contract Auditor and Solidity Programming Expert, who provide the model with detection techniques and relevant examples. The structured, scenario-based framework ensures targeted analysis and smooth information flow, using specific examples to guide the model toward identifying distinct vulnerabilities.

\subsection{Collaborative Decision-Making}
Collaborative decision-making process is another important component of action execution, ensuring that each step in the audit process benefits from the combined insights of multiple specialized agents. 

\subsubsection{Collaborative Analysis}
Each subtask within the system relies on effective communication between two specialized agents. To facilitate meaningful progress, the system employs a conversation-driven control flow, determining agent engagement and response processing. This approach enables intuitive reasoning about complex workflows, encompassing agent actions and message exchanges.

\begin{figure*}[t]
    \centering
    \includegraphics[width=1.0\linewidth]{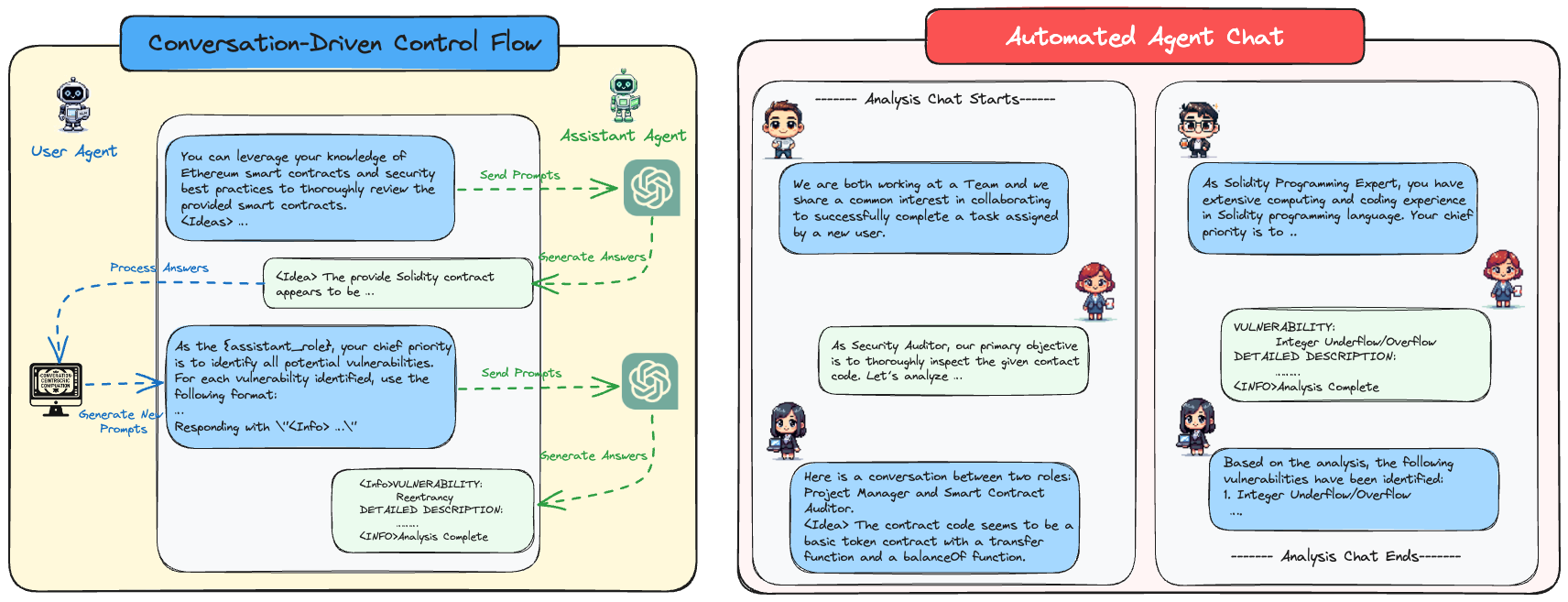}
    \caption{Collaborative Decision-Making between Two Agents}
    \label{fig:conversation_chat}
\end{figure*}

Figure \ref{fig:conversation_chat} illustrates the conversation-driven control flow and automated agent chat. The conversation-driven control flow demonstrates how task processes are executed between two agents through conversations. The process begins with the user agent sending a prompt to the assistant agent, which then generates a response via the language model API. The response is relayed back to the user agent, which generates a new prompt within the multi-conversation system. This new prompt is then sent to the assistant agent, initiating the next round of analysis.

The automated agent chat illustrates a simplified example of the smart contract analysis procedure. In this example, the system analyzes a contract potentially affected by an Arithmetic vulnerability. The process begins with the Project Manager initiating the audit task, setting the team's objective and initiating the discussion about the contract under review. As the analysis progresses, the Smart Contract Auditor and Project Manager contribute their expertise, sharing insights on the contract's purpose and structure. Finally, the Smart Contract Counselor summarizes the initial findings and provides a phase report. In the next phase, the Solidity Programming Expert provides a detailed code analysis, which the Auditor uses to identify potential Integer overflow/underflow vulnerabilities, offering a comprehensive description. The process concludes with the Counselor compiling a comprehensive audit report, summarizing all identified vulnerabilities and their potential impacts.

\subsubsection{Role Exchanges}
\begin{figure*}[ht]
    \centering
    \begin{subfigure}[b]{0.3\textwidth}
        \includegraphics[width=\linewidth]{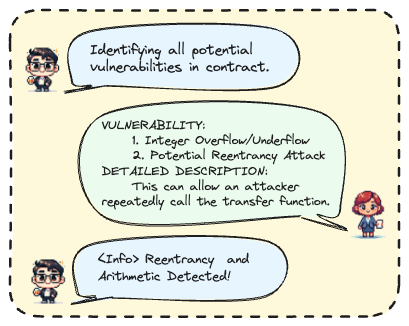}
        \caption{Simple QA}
        \label{fig:role1}
    \end{subfigure}
    \begin{subfigure}[b]{0.67\textwidth}
        \includegraphics[width=\linewidth]{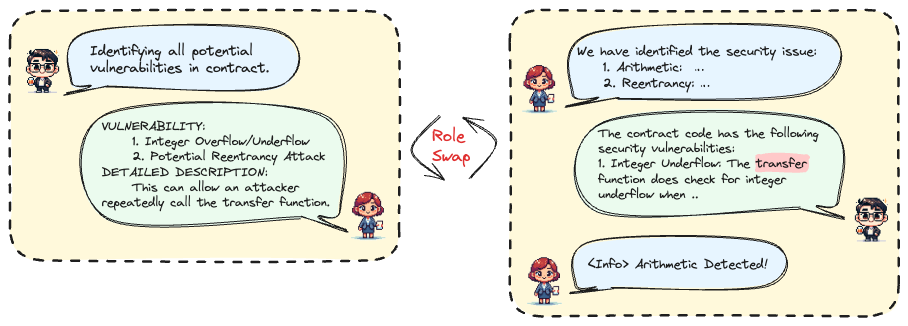}
        \caption{Role Exchanges QA}
        \label{fig:role2}
    \end{subfigure}
    \caption{Examples for Different Question-Answering}
    \label{fig:role-exchange}
\end{figure*}

Traditional LLM can sometimes produce inaccuracies or irrelevant information, especially in complex tasks like generating code insights.
For example, as illustrated in Figure \ref{fig:role1}, if the Smart Contract Auditor is instructed to review a vulnerable contract code previously identified with Arithmetic vulnerabilities, there's a risk of receiving misleading feedback. In such cases, the Auditor might erroneously flag the contract as vulnerable to both "Integer Overflow/Underflow" and "Reentrancy", incorrectly generating false positives for Reentrancy.
To address this issue, LLM-SmartAudit introduces a roles-swaping mechanism to enhance precision in vulnerability detection.

This innovative approach involves periodic roles exchanges between user and assistant agents. As shown in Figure \ref{fig:role2}, after the Auditor's initial analysis, roles are reversed, with the Solidity Programming Expert acting as the assistant agent. This role reversal enables the Expert to re-evaluate the contract from a fresh perspective, potentially identifying that what was initially perceived as an Arithmetic issue erroneously flagged as a Reentrancy vulnerability. 
The Auditor then reviews this revised analysis, making the final determination on the vulnerability classification.

In the final decision-making process, the system incorporates a consensus mechanism to assist the two agents. This mechanism facilitates cooperation between the user and assistant agents through multi-turn conversations, aiming to reach a consensus that ensures a well-informed and mutually agreed-upon final decision. This approach is crucial for ensuring agreement on the final audit results and other critical aspects, such as the contract's purpose and structure. However, reaching consensus may require multiple conversation rounds, potentially leading to extended deliberations. To streamline the process, the system limits discussions to a maximum of three rounds (where $n=3$), as shown in Figure \ref{task_queue}.

This section has outlined LLM-SmartAudit's framework, task queue, and collaborative decision-making process, establishing a comprehensive foundation for our investigation. The subsequent evaluation will assess LLM-SmartAudit's performance against leading traditional contract analysis tools.

\section{Evaluation}
\label{sec:experiments}
This section presents an evaluation of our system, comparing it with other smart contract detection tools.

\subsection{Research Questions}
We start our evaluation by posing the following research questions, focusing on the effectiveness of LLM-SmartAudit in detecting vulnerabilities in smart contracts:

\begin{itemize}
\item \textbf{RQ1: How does LLM-SmartAudit perform compared to leading traditional smart contract vulnerability detection tools in identifying specific vulnerability types?} 
This question evaluates LLM-SmartAudit's relative effectiveness in its BA mode against established tools, assessing its strengths and areas for potential improvement.
\item \textbf{RQ2: How do different strategies and GPT models affect LLM-SmartAudit's performance in detecting specific vulnerability types?}
This question examines the impact of analytical strategies (BA and TA) and different GPT models on the efficacy of smart contract vulnerability detection.
\item \textbf{RQ3: How does LLM-SmartAudit perform in real-world smart contracts?} 
This question evaluates LLM-SmartAudit's practical effectiveness in detecting vulnerabilities within real-world smart contract scenarios.
\item \textbf{RQ4: Can LLM-SmartAudit identify novel vulnerabilities overlooked in previous audit reports of the Real-world dataset?} 
This question explores LLM-SmartAudit's capability to detect vulnerabilities that were not identified in previous smart contract audits.
\end{itemize}

\begin{table*}[t]
  \centering
  \caption{Evaluation of Smart Contract Vulnerability Detection Tools — A Comparative Analysis}
  \scriptsize
  \label{tab:analysis_tools}
  \begin{threeparttable}
  \begin{tabular}{l | cccccccccc |c}
    \toprule
    \textbf{Tool} & \textbf{RE} & \textbf{IO} & \textbf{USE} & \textbf{UD} & \textbf{TOD} & \textbf{TM} & \textbf{RP} & \textbf{TX} & \textbf{USU} & \textbf{GS} & \textbf{Overall}\\
    \midrule
    Securify & 8 & - & 9 & - & 1 & - & - & - & - & - &18\% \\
    VeriSmart & - & 9 & - & - & -& - & - & - & - & - & 9\%\\
    Mythril-0.24.7 & 9 & 7 & 9 & 6 & - & 6 & 8 & 6 & 3 & - & 54\%\\
    Oyente & 7 & 9 & 5 & - & 2 & - & - & - & - & - & 23\%\\
    ConFuzzius & 9 & 7 & 9 & 1 & 2 & 8 & 2 & - & 4 & - & 42\% \\
    sFuzz & 6 & 6 & 6 & 5 & - & 1 & 6 & - & - & 3 &33\% \\
    Slither-0.10.0 & 9 & - & 8 & 7 & - & 8 & - & 8 & 6 & - & 46\%\\
    Conkas & 10 & 9 & 10 & - & 7 & 8 & - & - & - & - & 44\% \\
    GNNSCVD & 7 & - & - & - & - & - & 8 & - & - & - & 15\%\\
    Eth2Vec & 4 & 5 & - & - & - & 2 & - & - & - & 2 & 13\% \\
    BA (GPT-3.5-turbo) & 10 & 10 & 7 & 9 & 2 & 10 & 7 & 9 & 5 & 5 & \textbf{74\%}\\
    \bottomrule
  \end{tabular}
  \begin{tablenotes}
    \item Note: - indicates that a tool cannot detect this vulnerability type.
  \end{tablenotes}
  \end{threeparttable}
\end{table*}

\subsection{Experimental Settings}
To rigorously evaluate the effectiveness of our solution, we have established a transparent and reproducible experimental setup. This includes a multifaceted benchmarking dataset, evaluation criteria, and hardware configuration details.

\subsubsection{Benchmarking Dataset}
To comprehensively measure the system’s capabilities, robust evaluation datasets are essential. Zhang et al. \cite{zhang2023demystifying} categorize smart contract vulnerabilities into `machine-auditable' and `machine-unauditable' types. Traditional vulnerability detection tools can detect machine-auditable vulnerabilities, whereas machine-unauditable vulnerabilities require expert human intervention. In this study, machine-auditable vulnerabilities are termed `specific vulnerabilities', while machine-unauditable vulnerabilities are designated as `complex logic vulnerabilities'.

Based on this distinction, we created two datasets: the labeled dataset and the real-world dataset. Both datasets are publicly accessible via our repository.

\textbf{Labeled dataset} exclusively comprises specific vulnerabilities. The dataset encompasses ten types of vulnerabilities: Reentrancy (RE), Integer Overflow/Underflow (IO), Unchecked send (USE), Unsafe delegatecall (UD), Transaction Order Dependence (TOD), Time Manipulation (TM), Randomness Prediction (RP), Authorization Issue using `tx.origin' (TX), Unsafe Suicide (USU), and Gas Limitation (GL). The Labeled dataset consists of 110 annotated contract cases, categorized into 11 sub-datasets. Ten of these sub-datasets focus on individual specific vulnerability types, while the eleventh sub-dataset contains secure contracts.

\textbf{Real-world dataset} comprises both specific and complex logic vulnerabilities that have led to actual exploits. This dataset is derived from reputable Code4rena contests \cite{code4rena2024, zhang2023demystifying}, which attract global experts and companies to identify vulnerabilities in real-world smart contract projects. Participants receive financial compensation for their discoveries, lending credibility to the reports and confirming that the identified vulnerabilities genuinely reflect real-world attack scenarios. The Real-world dataset contains 102 projects and 6,454 contracts, encompassing 499 high-risk, 909 medium-risk, 1,420 low-risk, and 2,417 ground-level vulnerabilities. The severity classification of these vulnerabilities is based on their potential financial impact on the contract. This study primarily focuses on high-risk and medium-risk vulnerabilities.

\subsubsection{Evaluation Criteria}
In our investigation, the vulnerability detection process can be seen as a binary classification problem. The primary objective of the assessment tool is to accurately determine the presence or absence of specific vulnerabilities in a smart contract. 
This binary classification method simplifies the evaluation methodology and provides an effective measure of the tool's precision in vulnerability identification. 
The classification outcomes are categorized into four distinct groups: 

\begin{itemize}
  \item \textbf{True Positive (TP)}: The tool correctly identifies a vulnerability in a contract when one actually exists.
  \item \textbf{False Positive (FP)}: The tool incorrectly identifies a vulnerability in a contract when none exist.
  \item \textbf{False Negative (FN)}: The tool fails to identify a vulnerability when one actually exists.
  \item \textbf{True Negative (TN)}: The tool correctly identifies that a contract does not have a vulnerability when it does not.
\end{itemize}

To evaluate tool's performance, we use three key metrics: Precision, which is the ratio of true positive results to all positive results predicted by the tool (Precision = TP / (TP + FP)); Recall rates, which is the ratio of true positive results to all actual positive cases (Recall = TP / (TP + FN)); and F1-score, which is the harmonic mean of Precision and Recall (F1 = (2 * Precision * Recall) / (Precision + Recall)).  

\subsubsection{Hardware Configuration} This study utilized the gpt-3.5-turbo and gpt-4o versions of the GPT models, accessed via the OpenAI API\footnote{https://platform.openai.com/docs/concepts}. To enhance output stability from GPT, the default temperature was set to 0.2. All system evaluations were performed on an Aliyun-hosted Ubuntu 22.04 LTS machine, configured with an Intel(R) Core(TM) i5-13400 CPU and 32GB of RAM, ensuring consistent testing conditions.

\newmdenv[backgroundcolor=gray!20, linewidth=0.5pt, roundcorner=10pt, skipabove=\baselineskip]{custommdframed}

\begin{table*}[t]
  \centering
  \caption{Comparative Evaluation of Smart Contract Vulnerability Detection Across GPT-3.5 and GPT-4 Models}
  \label{tab:all_modes}
  \scriptsize
  \begin{threeparttable}
  \begin{tabular}{@{}llccccccccccccccc@{}}
      \toprule
      & \multicolumn{1}{l}{\multirow{2}{*}{\textbf{Type}}} & \multicolumn{5}{c}{\textbf{Zero-shot Prompt}} & \multicolumn{5}{c}{\textbf{BA Mode}} & \multicolumn{5}{c}{\textbf{TA Mode}} \\
      \cmidrule(lr){3-7} \cmidrule(lr){8-12} \cmidrule(l){13-17}
      & & TP & FN & FP & TN & F1-score & TP & FN & FP & TN & F1-score & TP & FN & FP & TN & F1-score \\
      \midrule 
      \multirow{10}{*}{\rotatebox{90}{GPT-3.5-turbo}} & RE & 10 & 0 & 3 & 7 & 87\% & 10 & 0 & 3 & 7 & 87\% & 10 & 0 & 1 & 9 & {95.2\%}(\textcolor{red}{$\uparrow$8.2\%}) \\
      & IO & 9 & 1 & 5 & 5 & 75\% & 10 & 0 & 2 & 8 & {90.9\%}(\textcolor{red}{$\uparrow$15.9\%}) & 10 & 0 & 1 & 9 & {95.2\%}(\textcolor{red}{$\uparrow$20.2\%}) \\
      & USE & 6 & 4  & 1 & 9 & 70.6\% & 7 & 3 & 0 & 10& 82.4\% (\textcolor{red}{$\uparrow$11.8\%}) & 10 & 0 & 1 & 9 & 95.2\% (\textcolor{red}{$\uparrow$24.6\%})\\ 
      & UD & 7 & 3 & 0 & 10 & 82.3\% & 9 & 1 & 0& 10& 94.7\% (\textcolor{red}{$\uparrow$12.4\%}) & 9 & 1 & 0 & 10 & 94.7\%\textcolor{red}{$\uparrow$12.4\%}\\ 
      & TOD & 0 & 10  & 0 & 10 & - & 2 & 8  & 0 & 10 & 33.3\% (\textcolor{red}{$\uparrow$33.3\%}) & 9 & 1 & 0 & 10 & 94.7\% (\textcolor{red}{$\uparrow$94.7\%})\\
      & TM &  9 & 1 & 0& 10 & 94.7\% & 10 & 0  & 0 & 10 & 100\% (\textcolor{red}{$\uparrow$5.3\%}) & 10 & 0 & 0 & 10 & 100\% (\textcolor{red}{$\uparrow$5.3\%}) \\
      & Rp & 7 & 3  & 0 & 10 & 82.4\%  & 7 & 3  & 0 & 10 & 82.4\%  & 10 & 0 & 0 & 10 & 100\% (\textcolor{red}{$\uparrow$17.6\%})\\ 
      & TX & 9 &  1 & 0 & 10 & 94.7\% & 9 & 1  & 0 & 10& 94.7\%  & 10 & 0 & 0 & 10 & 100\% (\textcolor{red}{$\uparrow$5.3\%}) \\ 
      & USU & 5 & 5  & 0 & 10 & 66.7\%& 5 & 5  & 0 & 10 & 66.7\%  & 7 & 3 & 2 & 8 & 73.7\% (\textcolor{red}{$\uparrow$6.6\%})\\
      & GL & 4 & 6 & 2 & 8 &50\% & 5 &  5 & 0 & 10 & 66.7\% (\textcolor{red}{$\uparrow$16.7\%}) & 9 & 1 & 0 & 10 & 94.7\% (\textcolor{red}{$\uparrow$44.7\%})\\
      \midrule 
      \multirow{2}{*}[-9ex]{ \rotatebox{90}{GPT-4o}} & RE &  10& 0  & 4 & 6 & 83.3\% & 10 & 0  & 1 & 9 & 95.2\% (\textcolor{red}{$\uparrow$11.9\%})& 10 & 0 & 1 & 9 & 95.2\% (\textcolor{red}{$\uparrow$11.9\%}) \\ 
      & IO & 10 & 0  & 2 & 8 & 90.9\% & 10 & 0  & 0 & 10 & 100\% (\textcolor{red}{$\uparrow$9.1\%}) & 10 & 0 & 0 & 10 & 100\% (\textcolor{red}{$\uparrow$9.1\%})\\ 
      & US & 8 & 2  & 0 & 10 & 88.9\% & 9 & 1  & 0 & 10 & 94.7\% (\textcolor{red}{$\uparrow$5.8\%}) & 9 & 1 & 0&10 & 94.7\% (\textcolor{red}{$\uparrow$5.8\%}) \\ 
      & UD & 9 &  1 & 1 & 9 & 90\% & 10 & 0 & 0 & 10 & 100\% (\textcolor{red}{$\uparrow$10\%}) & 10 & 0 & 0& 10 & 100\% (\textcolor{red}{$\uparrow$10\%}) \\ 
      & TOD & 3 & 7  & 0 & 10 & 46.2\% & 3 & 7 & 0 & 10 & 46.2\%  & 10 & 0 & 0 & 10 & 100\% (\textcolor{red}{$\uparrow$53.8\%})  \\
      & TM & 10 & 0  & 0 & 10 & 100\% & 10 & 0 & 0 & 10 & 100\%  & 10 & 0 & 0 & 10 & 100\% \\
      & RP & 10 & 0  &  0& 10 & 100\% & 10 & 0 & 0 & 10 & 100\%  & 10 & 0 & 0 & 10 & 100\%  \\ 
      & TX & 10 & 0  & 0 & 10 & 100\% & 10 & 0 & 0 & 10 & 100\% & 10 & 0 & 0 & 10 & 100\%  \\ 
      & USU & 9 & 1  & 0 & 10 & 95\% & 10 & 0  &  0 & 10  & 100\% (\textcolor{red}{$\uparrow$5\%}) & 10 & 0 & 0 & 10 & 100\% (\textcolor{red}{$\uparrow$5\%}) \\
      & GL & 5 & 5  & 1 & 9 & 62.5\% & 6 & 4 & 0 & 10 & 75\% (\textcolor{red}{$\uparrow$12.5\%}) & 9 & 1 & 0 & 10 & 94.7\% (\textcolor{red}{$\uparrow$32.2\%}) \\
      \bottomrule
  \end{tabular}
  \begin{tablenotes}
      \item Note: \textcolor{red}{$\uparrow$} indicates improved values compared to the Zero-shot prompt.
  \end{tablenotes}
\end{threeparttable}
\end{table*}

\subsection{Experimental Results}
To answer the research questions posed earlier, we present the experimental results.
\subsubsection{RQ1: Comparative Evaluation with Leading Traditional Tools}
To address RQ1, we evaluated the effectiveness of LLM-SmartAudit in its BA mode, utilizing the GPT-3.5 model. The study evaluates the tool's capability to identify 10 specific vulnerabilities types in smart contracts, comparing its performance against leading traditional tools in the domain. The comparative analysis included ten prominent smart contract analysis tools across various categories: 
formal verification (Securify \cite{tsankov2018securify}, VeriSmart \cite{so2020verismart}), symbolic execution (Mythirl \cite{Mythriltool}, Oyente \cite{atzei2017survey}), fuzzing (ConFuzzius  \cite{torres2021confuzzius}, sFuzz \cite{nguyen2020sfuzz}), intermediate representation (IR) analysis (Slither \cite{feist2019slither}, Conkas \cite{Veloso2021}), and machine learning approaches (e.g., GNNSCVD \cite{zhuang2021smart}, Eth2Vec \cite{ashizawa2021eth2vec}). Table \ref{tab:analysis_tools} presents detailed performance metrics for each analyzed tool, specifically focusing on False Positives (FPs) and recall rates.

The results demonstrate that BA mode achieves the highest overall recall rates of 74\%, significantly outperforming all other evaluated tools. Mythril ranks second with an overall recall rates of 54\%, followed by Slither at 46\%. Notably, BA mode exhibited exceptional versatility, detecting all ten specific types of vulnerabilities with competitive overall recall rates. In contrast, traditional tools typically excel in detecting specific types of vulnerabilities but often struggle with others. For instance, while Conkas performs exceptionally well in identifying RE, IO, and USE vulnerabilities, it fails to detect several other vulnerability types.

However, our method shows limitations in detecting certain vulnerabilities, like TOD, USU, and GS. While LLM-SmartAudit in BA mode demonstrates superior overall performance and versatility in vulnerability identification, these limitations indicate areas for potential improvement. These limitations are primarily attributable to the inherent capabilities of the underlying model. For instance, TOD are often highly context-dependent and may require a deeper understanding of the specific sequence of transactions and their interactions, which might not be fully encapsulated by the GPT model. These findings underscore the need for additional strategies or more powerful GPT models to enhance detection accuracy.

\begin{custommdframed}
\textbf{Answer to RQ1:} Our method (BA mode) demonstrates superior overall performance in detecting a diverse range of smart contract vulnerabilities when compared to leading traditional tools. The method's capacity to maintain high recall rates across various vulnerability types underscores the potential of LLMs in smart contract auditing.
\end{custommdframed}

\subsubsection{RQ2: Performance of LLM-SmartAudit Across Different Strategies}
The analysis conducted for RQ1 revealed specific limitations of LLM-SmartAudit in BA mode, particularly in detecting TOD vulnerabilities.  To address these limitations and answer RQ2, LLM-SmartAudit was adapated to the TA mode, and the LLM was upgraded to the more advanced GPT-4 model. Additionally, we introduced a baseline comparison using a zero-shot prompt approach. Table \ref{tab:all_modes} presents the comprehensive performance metrics for each strategy, including TP, FP, FN, TN, and F1-score. 

The results indicate that both BA and TA modes significantly enhance the detection capabilities of GPT model compared to the zero-shot prompt baseline. The BA mode effectively reduces FPs across both models, highlighting its efficacy in refining the decision-making process and mitigating uncertainties in LLM outputs.
TA mode exhibits superior performance across both models, particularly in detecting complex vulnerabilities like Timestamp Dependency and Gas Limitation, achieving high or perfect F1-scores. These results underscore the efficacy of TA mode in focusing model attention and enhancing detection precision.

Furthermore, the analysis reveals that GPT-4 consistently outperforms GPT-3.5 across all modes, reflecting advancements in model architecture and training datasets. This performance improvement is particularly evident in the higher F1-scores achieved for critical vulnerabilities, including Integer Overflow, Timestamp Dependency, and Gas Limitation.

Importantly, TA mode utilizing GPT-3.5 can outperform zero-shot prompts with GPT-4 in specific scenarios such as detecting TOD and Gas Limitation vulnerabilities. These findings suggest that the strategic application of LLM agents can surpass the inherent limitations of different GPT versions, offering a more cost-effective solution while maintaining detection efficacy.

\begin{custommdframed}
  \textbf{Answer to RQ2}: This analysis demonstrates that while both GPT models are effective in detecting smart contract vulnerabilities, their performance is significantly enhanced by applying specialized strategies (BA and TA modes). Moreover, the excellent performance of TA mode suggests that strategic application can enable weaker model to achieve a level of effectiveness comparable to more advanced model.
\end{custommdframed}

\begin{table}[!ht]
    \centering
    \caption{Performance Comparison of Different Detection Methods on Real-World Dataset}
    \scriptsize
    \label{tab:contest_dataset}
    \begin{threeparttable}
        \begin{tabular}{l *{4}{c}}
            \toprule
            \multirow{3}{*}{\textbf{Tool}} & \multicolumn{2}{c}{\textbf{Specific Type}} & \multicolumn{2}{c}{\textbf{Complex Logic Type}} \\
            \cmidrule(lr){2-3} \cmidrule(lr){4-5}
            & \textbf{TP} & \textbf{Recall (\%)} & \textbf{TP} & \textbf{Recall (\%)} \\
            \midrule
            Slither-0.10.0 & 0 & 0.00 & 0 & 0.00 \\
            Mythril-0.24.7 & 0 & 0.00 & 0 & 0.00 \\
            Conkas         & 0 & 0.00 & 0 & 0.00 \\
            \midrule
            BA mode        & 92 & 30.26 & 110 & 9.96 \\
            TA mode        & \textbf{147} & \textbf{48.36} & \textbf{525} & \textbf{47.55} \\
            \bottomrule
        \end{tabular}
        \begin{tablenotes}[flushleft]
            \item \textit{Note:} The real-world dataset contains 304 specific vulnerabilities and 1,104 complex logic vulnerabilities.
        \end{tablenotes}
    \end{threeparttable}
\end{table}

\subsubsection{RQ3: Evaluation Using Real-world Dataset}
To address RQ3, we present a comprehensive evaluation of LLM-SmartAudit system using the Real-world dataset. The LLM-SmartAudit system design allocates approximately 1,000 tokens for prompt engineering, excluding the context of code snippets.
Given GPT-3.5's default token limit of 4,096, only 3,000 tokens remain available for contract content within the prompt. For the Labeled dataset, the analyzed contracts are less than 3,000 tokens in length. However, with the Real-world datasets, 533 contracts exceed 3,000 tokens in length. Consequently, only GPT-4 was utilized for this evaluation due to its higher token support of up to 128,000 tokens in context. The evaluation focuses on two primary metrics: TPs and recall, with results compared against audit reports\footnote{https://github.com/ZhangZhuoSJTU/Web3Bugs/tree/main/reports}. Table \ref{tab:contest_dataset} presents the evaluation results.

The results indicate that the TA mode consistently outperforms other tools in detecting both specific type and complex logic type vulnerabilities. The BA mode shows moderate effectiveness for specific type vulnerabilities but struggles with complex logic type vulnerabilities. The traditional tools (Conkas, Slither-0.10.0 and Mythril-0.24.7)  demonstrated no effectiveness in detecting the vulnerabilities reported in this dataset, indicating potential limitations in their capability to identify these particular types of vulnerabilities.

The superior performance of TA mode can be primarily attributed to its carefully designed set of 40 specific scenarios. These specific scenarios effectively narrow the operational context for the LLM, enhancing its focus. This focused approach aligns with how LLMs process information, enabling them to concentrate on relevant patterns and structures within the smart contract code that are indicative of particular vulnerabilities. Consequently, TA mode demonstrates higher efficacy compared to the broader scanning method employed by BA mode, especially for complex logic vulnerabilities.

While earlier works such as David et al. \cite{david2023you} and GPTScan \cite{sun2024gptscan} demonstrated the potential of GPT-based vulnerability detection tools, our research significantly broadens the scope and improves upon their findings. However, due to the unavailability of aforementioned tools and insufficient information for replication,  this study relies on their published statistics for comparison. David et al. reported a recall rate of 39.73\%, detecting 58 out of 146 vulnerabilities using a combination of Claude and GPT. GPTScan analyzed 232 vulnerabilities across 72 projects, correctly identifying 40 true positives. In contrast, our study substantially broadens the scope, examining 1,408 vulnerabilities across 102 projects. This expanded scope allows for a more comprehensive assessment of GPT-based tools' capabilities in real-world scenarios. Our results show a notable improvement, with 672 out of 1,408 vulnerabilities detected, yielding an overall recall rate of 47.73\%. 

\begin{custommdframed}
  \textbf{Answer to RQ3}: These findings highlight the superiority of BA and TA modes over traditional detecting tools. The enhanced performance of TA mode suggests that carefully designed, targeted scenarios can significantly improve LLMs's in identifying complex vulnerabilities in smart contracts. The results also underscore the necessity for continued improvement in vulnerability detection tools, particularly for complex logic vulnerabilities.
\end{custommdframed}

\subsubsection{RQ4: Newly Discovered Vulnerabilities} In response to RQ4, our methods successfully identified 11 vulnerabilities across 4 different types that were not detected in the audit reports from real-world datasets. These findings have been submitted to the Code4rena community for verification. Table \ref{table:issues} provides the list of these newly discovered vulnerabilities.

\begin{table}[ht]
\caption{Summary of Newly Discovered Vulnerabilities}
\label{table:issues}
\scriptsize
\begin{tabularx}{\columnwidth}{>{\raggedright\arraybackslash}p{2cm} X}
\toprule
\textbf{Vulnerability} & \textbf{Affected Locations} \\
\midrule
Unlimited Token Approval & 
    \texttt{supplyTokenTo()} in \texttt{SushiYieldSource.sol}; 
    \texttt{.approve()} in \texttt{NFTXStakingZap.sol}; 
    \texttt{\_par.approve()} in \texttt{PARMinerV2.sol} \\
\midrule
Insufficient Input Validation & 
    \texttt{setTransferRatio()} in \texttt{sYETIToken.sol} \\
\midrule
Improper Partial Withdrawals & 
    \texttt{withdraw()} in \texttt{yVault.sol}; 
    \texttt{processWithdraw()} in \texttt{synthVault.sol} \\
\midrule
Unchecked External Calls & 
    \texttt{earn()}, \texttt{withdraw()} in \texttt{yVault.sol}; 
    \texttt{vault.addValue()}, \texttt{getUnifiedAssets()} in \texttt{IndexTemplate.sol}; 
    \texttt{collectEarnings()}, \texttt{\_push()}, \texttt{\_pullUniV3Nft()} in \texttt{UniV3Vault.sol} \\
\bottomrule
\end{tabularx}
\end{table}

One notable example is the `Unlimited Token Approval' vulnerability found in the `SushiYieldSource.sol' contract. In function `supplyTokenTo', the contract calls `sushiAddr. approve(address(sushiBar), amount);' which approves the SushiBar contract to spend the specified `amount' of tokens. If the `amount' is significantly larger than what is necessary for the current operation, it can lead to a situation where the SushiBar contract has excessive approval to spend tokens on behalf of the user. This can be exploited if the SushiBar contract is compromised or behaves unexpectedly, allowing an attacker to drain tokens from the user's account.

\begin{custommdframed}
  \textbf{Answer to RQ4:} 
    Our methods identified 11 new vulnerabilities that were not presented in the Real-world dataset audit reports. This discovery highlights the capability of our method to identify subtle yet critical vulnerabilities that may have been overlooked in traditional auditing.
\end{custommdframed}

\subsection{Related Work}
\label{sec:related}
The application of LLMs in programming is well-established, yet their efficacy in domain-specific languages (DSLs) like Solidity remains an emerging area of research. Recent studies have begun to explore the potential of general-purpose LLMs such as GPT in the domain of smart contract security analysis. 

David et al. \cite{david2023you} examined the efficacy of LLMs, like GPT-4 and Claude, in auditing DeFi smart contract security. Their study employed a binary classification approach, asking the LLMs to determine whether a contract is vulnerable. Although GPT-4 and Claude demonstrated high true positive rates,  they also exhibited significant false positive rates. The researchers highlighted the substantial evaluation cost, approximately 2,000 USD, for analyzing 52 DeFi attacks.
Chen et al. \cite{chen2023chatgpt} performed a comparative analysis of GPT's smart contract vulnerability detection capabilities against established tools. Their results revealed varying GPT effectiveness across common vulnerability types, encompassing 8 types compared to the 10 in our study. Sun et al. \cite{sun2024gptscan} evaluated GPT's function vulnerability matching using a binary response format ('Yes' or 'No') for predefined scenarios. They also highlighted potential false positives due to GPT's inherent limitations. While their study found no significant improvements with GPT-4, our research and others have demonstrated GPT-4's enhanced detection capabilities.

In addition to SOTA commercial products, open-source alternatives have been considered for smart contract analysis. Shou et al. \cite{shou2024llm4fuzz} integrates Llama-2 model into the process of fuzzing to detect vulnerabilities in smart contracts, aiming to address inefficiencies in traditional fuzzing methods. However, this approach's efficacy depends on LLMs' accurate and nuanced understanding of smart contracts, and it confronts challenges in complexity, cost, and dependence on static analysis. Sun et al. \cite{sun2024llm4vuln} explored open-source tools such as Mixtral and CodeLlama against GPT-4 for detecting smart contract vulnerabilities. They discovered that GPT-4, leveraging its advanced Assistants’ functionalities to effectively utilize enhanced knowledge and structured prompts, significantly outperformed Mixtral and CodeLlama. However, the assessment was limited to demonstrations from the Replicate website, potentially not fully representing these LLMs' capabilities.

\section{Discussions}
\label{sec:discussion}

\subsection{Summary of Findings}
Based on the above evaluation results, we have derived the following key findings:

\begin{itemize}
    \item \textbf{Collaborative Multi-Agents}: Our research reveals that utilizing multiple LLM-based agents provides a thorough approach to smart contract security. Specialized agents with role-specific instructions conduct deeper analyses in their respective areas compared to general LLM knowledge. The synergy among diverse agents emulates the operations of a professional smart contract auditing firm, significantly enhancing the reliability and comprehensiveness of the audits.

    \item \textbf{Advanced Detection}: LLM-SmartAudit outperforms leading traditional tools in identifying a wide range of smart contract vulnerabilities, demonstrating the potential of LLMs in this field. Both LLM-SmartAudit's BA and TA modes enhance vulnerability detection compared to zero-shot prompts to GPT models. Notably, the TA mode's scenario-based strategy significantly improves the LLMs' ability to detect complex vulnerabilities.
    
    \item \textbf{Flexibility and Adaptability}: LLM-SmartAudit's combined BA and TA modes provide the adaptability needed to integrate emerging vulnerability patterns. TA excels at detecting known vulnerabilities, while BA is adept at discovering unknown ones. Newly identified vulnerabilities can be fed back into the TA mode as new detector types, enhancing its detection capabilities to keep pace with evolving security threats in the smart contract landscape. Additionally, the system supports not only GPT-3.5-turbo and GPT-4, but also various other models, thereby increasing its versatility.
    \item \textbf{Cost Effectiveness}: Conventional smart contract auditing firms often charge high fees, with basic audits ranging from 500 USD for a simple review to several thousand dollars for comprehensive assessments. In contrast, LLM-SmartAudit offers a fully automatic solution with a maximum cost of just 1 USD per contract. This significant cost reduction, combined with its ability to uncover a variety of complex vulnerabilities that human experts might overlook, positions LLM-SmartAudit as a highly efficient and economical alternative in the field of smart contract auditing.
\end{itemize}

\subsection{Threats of Validity}
Our proposed system has the following potential limitations:

\begin{itemize}
  \item \textit{Model Dependence}: The effectiveness of LLM-SmartAudit is closely tied to the capabilities of LLMs. Current GPT models excel in data processing speed, inter-network responses, and token usage efficiency. Although local LLMs provide an alternative, they require significant resources, making them a substantial investment, particularly for smaller teams. Given these resource constraints, API-based solutions are the most practical approach.
  \item \textit{Evolving Vulnerability Landscape}: Our TA mode currently covers 40 scenarios, effectively identifying numerous vulnerability types previously found by human experts. However, this approach may not be exhaustive. Complex vulnerabilities, particularly those arising from emerging or previously unreported issues, remain challenging to detect. 
  While powerful, our method's reliance on static analysis presents inherent limitations in identifying dynamic vulnerabilities.
\end{itemize}

This section has provided a comprehensive overview of the LLM-SmartAudit system, highlighting its strengths and limitations. These insights form our study, offering a nuanced understanding of the potential and challenges of using LLMs in smart contract analysis.

\section{Conclusions}
\label{sec:conclusion}
In this paper, we introduced the LLM-SmartAudit system, a novel framework for automatically detecting vulnerabilities in smart contracts. Our system employs a multi-agent conversational approach to simulate a role-based virtual auditing organization. We demonstrated the potential of large language models, particularly GPT models, in identifying a wide array of vulnerabilities in smart contracts. Through comprehensive evaluation, we compared our system with traditional tools, highlighting its effectiveness. 
LLM-SmartAudit emerges as a robust solution for enhancing smart contract security. It offers a more efficient and effective approach to vulnerability detection while also opening new avenues for future research in this field. Future directions for LLM-SmartAudit include integrating more advanced language models, expanding its capabilities to handle a broader range of vulnerability types, and refining its algorithms to improve detection accuracy.

\bibliographystyle{ACM-Reference-Format}
\bibliography{references}

\end{document}